\documentclass[11pt]{article}
\usepackage{epsfig}
\usepackage{float}
\usepackage{graphicx}
\usepackage{cite}
\usepackage{amsmath}
\usepackage{amssymb}
\usepackage{amsfonts}
\usepackage{appendix}
\usepackage[utf8]{inputenc}
\usepackage[left=25mm,top=25mm,bottom=25mm,right=25mm]{geometry}
\usepackage{multirow}
\usepackage{hhline}
\usepackage{booktabs}
\usepackage{amssymb}
\usepackage{hyperref}
\usepackage{xcolor}
\usepackage{graphicx}
\usepackage{dcolumn}
\usepackage{bm}
\usepackage{pifont}
\begin{document}
\bigskip

\newcommand{\be}{\begin{equation}}
\newcommand{\en}{\end{equation}}
\newcommand{\bea}{\begin{eqnarray}}
\newcommand{\ena}{\end{eqnarray}}
\newcommand{\noi}{\noindent}
\newcommand{\ra}{\rightarrow}
\newcommand{\bib}{\bibitem}
\newcommand{\bff}{\begin{figure}}
\newcommand{\eff}{\end{figure}}

\begin{center}
{\Large \bf Regular charged black holes, energy conditions and quasinormal modes.}
\end{center}
\hspace{0.0 cm}


\centerline{
Leonardo Balart$^{\triangle}$, 
Grigoris Panotopoulos$^{\spadesuit}$, 
\'Angel Rinc{\'o}n$^{\clubsuit}$,}

\vskip 0.7cm

\centerline{Departamento de Ciencias F{\'i}sicas, Universidad de la Frontera,} \centerline{Casilla 54-D, 4811186 Temuco, Chile.}
\centerline{$^{\triangle}$email:
\href{mailto:leonardo.balart@ufrontera.cl}
{\nolinkurl{leonardo.balart@ufrontera.cl}}
}

\vskip 0.3cm

\centerline{Departamento de Ciencias F{\'i}sicas, Universidad de la Frontera,} \centerline{Casilla 54-D, 4811186 Temuco, Chile.}
\centerline{$^{\spadesuit}$email:
\href{mailto:grigorios.panotopoulos@ufrontera.cl}
{\nolinkurl{grigorios.panotopoulos@ufrontera.cl}}
}

\vskip 0.3cm

\centerline{Departamento de Física Aplicada, Universidad de Alicante,}
\centerline{Campus de San Vicente del Raspeig, E-03690 Alicante, Spain.}
\centerline{Sede Esmeralda, Universidad de Tarapacá,}
\centerline{ Avda. Luis Emilio Recabarren 2477, Iquique, Chile.}
\centerline{$^{\clubsuit}$email:
\href{mailto:angel.rincon@ua.es}
{\nolinkurl{angel.rincon@ua.es}}
}

\hspace{0.3 cm}

\begin{abstract}
We discuss energy conditions and quasinormal modes for scalar perturbations of regular charged black holes within the framework of General Relativity coupled to non-linear electrodynamics. The frequencies are computed numerically adopting the WKB method, while in the eikonal limit an analytic expression for the spectra is obtained. The impact of the electric charge, the angular degree, and the overtone number on the spectra is investigated in detail. We find that all frequencies are characterized by a negative imaginary part, and that each type of energy conditions imply a different quasinormal spectrum.
\end{abstract}

\hspace{0.8cm}



\section{Introduction}
\label{intro}

Realistic black holes are not isolated in Nature. Instead, they are in constant interaction with their environment.
When a black hole is perturbed, the geometry of space-time undergoes damped oscillations. How a system responds to 
small perturbations has always been important in physics. The work of \cite{regge} marked the birth of
black hole perturbations, it was later extended by \cite{zerilli1,zerilli2,zerilli3,moncrief,teukolsky}, while the state-of-the art in black hole perturbations is 
summarized in the comprehensive review of Chandrasekhar's monograph \cite{monograph}. Quasi-normal (QN) frequencies are complex numbers, with a non-vanishing imaginary part,
that encode the information on how black holes relax after the perturbation has been 
applied. They depend on the geometry itself as well as the type of the perturbation (scalar, Dirac, vector (electromagnetic), tensor (gravitational)). As 
they do not depend on the initial conditions, QN modes (QNMs) carry unique information about black holes. Black hole perturbation
theory and QNMs of black holes are relevant during the ringdown phase of binaries, in which after the merging of two black 
holes a new, distorted object is formed, while at the same time the geometry of space-time undergoes damped 
oscillations due to the emission of gravitational waves.

\smallskip

Given the interest in gravitational wave astronomy and in QNMs of black holes, it would be interesting to see what kind
of QN spectra are expected from regular electrically charged black holes in non-linear electrodynamics. In the present work we propose to compute QNMs for scalar perturbations of black hole solutions with a net electric charge within four distinct non-linear electrodynamics (NLE) models fulfilling different energy conditions.

\smallskip

There are certain static black hole solutions that have an event horizon and whose metric and curvature invariants $R$, $R_{\mu\nu}R^{\mu\nu}$, $R_{\kappa\lambda\mu\nu}R^{\kappa\lambda\mu\nu}$ do not present singularities in the whole interval of the radial coordinate $r$. Such solutions are known as regular (or non-singular) black holes. Some of these regular solutions~\cite{Dymnikova:1992ux, Nicolini:2005vd, Hayward:2005gi, Xiang:2013sza, Ghosh:2014pba, Frolov:2016pav, Frolov:2017rjz} have no electric or magnetic charge, they depend on the mass of the black hole and some extra parameter and asymptotically behave as the Schwarzschild solution.
Other solutions have electric (or magnetic) charge, but in this group are those that do not behave like the Reissner-Nordstrom solution in the limit of weak fields, for example the Bardeen solution~\cite{Bardeen:1968}, which can be obtained from a nonlinear electrodynamics of a magnetic source~\cite{Ayon-Beato:2000mjt} or of a electric source~\cite{Rodrigues:2018bdc}. The Hayward solution given in Ref.~\cite{Hayward:2005gi} can also be considered as such a charged solution of this type~\cite{Balart:2014jia} (see also other examples given in this same reference).
A large number of regular black hole solutions have been found using nonlinear electromagnetic sources, which in the weak field limit behave like the Reissner-Nordstrom solution, for example~\cite{AyonBeato:1998ub, AyonBeato:1999rg, Bronnikov:2000vy, Dymnikova:2004zc, Balart:2014cga}.  For recent literature about regular black holes see for instance \cite{Javed:2022rrs,Ovgun:2019wej,Jusufi:2018jof,Contreras:2017eza,Melgarejo:2020mso,Contreras:2020fcj}.

The energy conditions are the constraints for the energy-momentum tensor $T^{\mu\nu}$ of a theory of gravity~\cite{Hawking:1973uf}. 
The standard acceptable conditions assumed for the energy-momentum tensor are:
weak energy condition (WEC), dominant energy condition (DEC), null energy condition (NEC), and strong energy condition (SEC) (see Refs.~\cite{Hawking:1973uf,Wald:1984rg,Frolov:1998wf}).
If $\xi_\mu$ and $k_\mu$ are arbitrary timelike and null vectors, respectively, then the conditions for the energy-momentum tensor are expressed with the following inequalities

\begin{equation}
T^{\mu\nu} \xi_\mu \xi_\nu \geq 0     \,\,\,\,\,\, \mbox{(WEC)}.
\label{wec}
\end{equation}
\begin{equation}
T^{\mu\nu} \xi_\mu \xi_\nu \geq 0   \mbox{ and } T^{\mu\nu} \xi_\mu \mbox{ is a non-spacelike vector} \,\,\,\,\,\, \mbox{(DEC)}.
\label{dec}
\end{equation}
\begin{equation}
T^{\mu\nu} k_\mu k_\nu \geq 0 \,\,\,\,\,\, \mbox{(NEC)}.
\label{nec}
\end{equation}
\begin{equation}
T^{\mu\nu}\xi_\mu \xi_\nu - \frac{1}{2} T^{\mu}_{\,\,\,\mu} \xi^\nu \xi_\nu\geq 0 \,\,\,\,\,\, \mbox{(SEC)}.
\label{sec}
\end{equation}


The above inequalities imply for a perfect fluid the energy conditions
\begin{equation}
\rho \geq 0,
\end{equation}
\begin{equation}
\rho \pm p \geq 0,
\end{equation}
\begin{equation}
\rho + 3 p \geq 0,
\end{equation}
where $p,\rho$ are the pressure and the energy density of the fluid, respectively.

\smallskip

The regular black holes violate the SEC somewhere inside the event horizon~\cite{Zaslavskii:2010qz}. However, a regular black hole could satisfy the NEC or the WEC or the DEC everywhere. Those regular black holes that satisfy the WEC necessarily have asymptotically de Sitter behavior for $r \rightarrow 0$~\cite{Dymnikova:2001fb}.
The WEC together with the symmetry $T_{\,\,\,0}^0 = T_{\,\,\,1}^1$ for the energy-momentum tensor have been used to obtain new solutions of regular black holes~\cite{Balart:2014jia, Rodrigues:2017yry} and without it having been a requirement that these solutions also satisfy the DEC everywhere.


Regular black hole solutions can be classified either by the energy conditions that they satisfy or by the way in which they do not satisfy these energy conditions.

\smallskip

In the present article our work is organized as follows: After this introductory section, in section 2 we summarize the field equations for Einstein gravity coupled to non-linear Electrodynamics. In the third section we discuss energy conditions, while in section 4 the stability of the BH solutions analyzed here is discussed. Finally, in the fifth section we compute the QN spectra, and we conclude our work in section 6. We adopt the mostly positive metric signature $(-,+,+,+)$, and we work in geometrical units where $G=1=c$.

\section{Regular charged black holes}
\label{RBH}

The solutions of charged regular black holes can be obtained from the following action of general relativity coupled with a nonlinear electrodynamics model
\begin{equation}
S =  \int d^4 x  \sqrt{-g} \left(\frac{R}{16 \pi} - \mathcal{L}(\mathcal{F})\right)    \, ,
\label{action}
\end{equation}
where $g$ is the determinant of the metric tensor, $R$ is the Ricci scalar, and $\mathcal{L}(\mathcal{F})$ is the Lagrangian that depends on the Lorentz invariant $\mathcal{F} = F^{\mu\nu}F_{\mu\nu}/4$. In these theories the electrodynamic model is expressed by a Hamiltonian function $\mathcal{H}(\mathcal{P})$ and fields $P^{\mu\nu}$ instead of $\mathcal{L}(\mathcal{ F})$ and the fields of the electromagnetic tensor $F^{\mu\nu}$, but which are related by the Legendre transformation
$\mathcal{L}= P_{\mu\nu} F^{\mu\nu} - \mathcal{H}$.

Here we adopt the following most general form of a static line element with spherical symmetry
\begin{equation}
ds^2 =   -f(r) dt^2 + f(r)^{-1}dr^2 + r^2(d\theta^2 + \sin^2 \theta d^2\phi)      \, ,
\label{element-g}
\end{equation}
where the metric function is given by
\begin{equation}
f(r) = 1 - \frac{2 m(r)}{r}     \, ,
\label{metric-g}
\end{equation}
and $m(r)$ is the mass function that depends on the mass $M$, the electric charge $q$ of the black hole as well as the radial coordinate $r$.

Computing the variation of the action~(\ref{action}) allows us to obtain the Einstein field equations
\begin{equation}
G_{\mu \nu}=R_{\mu \nu}-\frac{1}{2} R g_{\mu \nu}=8 \pi T_{\mu \nu} \, ,
\label{Guv}
\end{equation}
and
\begin{equation}
\nabla_{\mu} \left( F^{\mu \nu} \frac{d \mathcal{L}}{d \mathcal{F}}\right)=0 \, .
\label{E L_F}
\end{equation}
Here 
\begin{equation}
T_{\mu \nu}=\frac{1}{4 \pi} \left(g_{\mu \nu} \mathcal{L}(\mathcal{F}) -  F_{\mu \alpha} F_{\nu}^{\, \alpha}  \frac{d \mathcal{L}}{d \mathcal{F}} \right)  \, .
\label{tensor-g}
\end{equation}

From these last three equations we can obtain a general expression for the electric field $\mathbf{E} = E(r)\hat{r}$ in terms of the mass function
\begin{equation}
E(r) =\frac{r}{2q} \left(\frac{2}{r}\frac{dm(r)}{dr} - \frac{d^2m(r)}{dr^2} \right)   \, .
\label{G-E-Field}
\end{equation}


\section{Energy conditions and regular charged black holes}
\label{EC}
We can conveniently express the energy conditions given above, if we note that from Eqs.~(\ref{element-g}) and (\ref{metric-g}) we obtain
\begin{equation}
G^0_{\,\,0} = G^1_{\,\,1} = \frac{1}{r^2}\left(-1 + f(r) + \frac{df(r)}{dr} \right) = 
- 2 \frac{dm(r)}{dr}  \ , 
\label{}
\end{equation}
\begin{equation}
G^2_{\,\,2} = G^3_{\,\,3} = \frac{1}{r}\frac{df(r)}{dr} + \frac{1}{2}\frac{d^2f(r)}{dr^2} = 
- \frac{1}{r} \frac{d^2m(r)}{dr^2}   \ , 
\label{}
\end{equation}
and therefore the field equations~(\ref{Guv}) imply for the stress-energy tensor
\begin{equation}
T^0_{\,\,0} = T^1_{\,\,1} = -\frac{2}{8 \pi r^2}\frac{dm(r)}{dr}  \ , 
\label{}
\end{equation}
\begin{equation}
T^2_{\,\,2} = T^3_{\,\,3} = -\frac{1}{8 \pi r}\frac{d^2m(r)}{dr^2}  \ .
\label{}
\end{equation}
Thus we can write that the WEC requires the following inequalities to be satisfied for all $r$
\begin{equation}
\frac{1}{r^2}\frac{dm(r)}{dr} \geq 0 \ , 
\label{EC-1}
\end{equation}
and
\begin{equation}
 \frac{2}{r}  \frac{d m(r)}{dr}  - \frac{d^2 m(r)}{dr^2}  \geq 0  \, .
\label{EC-2}
\end{equation}
where the first one is equivalent to the condition $\rho \geq 0$ (or $-T_0^0 \geq 0$), while the other one is equivalent to the condition $p + \rho \geq 0$ (or $-T_0^0+T_2^2 \geq 0$).

\smallskip

Meanwhile, the DEC requires that the two previous inequalities be satisfied along with the following       inequality
\begin{equation}
\frac{2}{r}  \frac{d m(r)}{dr} +  \frac{d^2 m(r)}{dr^2}  \geq 0 \, .
\label{EC-3}
\end{equation}
The NEC requires that only inequality~(\ref{EC-2}) be satisfied.

Note that the DEC implies the WEC and in turn the WEC implies the NEC. Also note that from the expression~(\ref{EC-2}) obtained for the electric field, if the WEC (or NEC) is violated on some interval, then on this same interval the electric field is negative (where the electric charge is $|q|$).


Based on the energy conditions mentioned above, we can classify static regular black holes into three types:

\vspace{0.3cm}
\noindent 1. Those who satisfy the DEC everywhere and therefore also the WEC (and the NEC).

\vspace{0.3cm}
\noindent 2. Those who satisfy the WEC everywhere, but do not satisfy the DEC somewhere.
 
 \vspace{0.3cm}
\noindent 3. Those who do not satisfy the WEC somewhere in $r$. In terms of the inequalities given above, we see that there are two different ways of not satisfying it. In both cases inequality~(\ref{EC-2}) is not fulfilled, but in one case inequality~(\ref{EC-3}) is fulfilled and in the other case is not.

\vspace{0.3cm}
Below, we consider an example of a regular black hole of each type, we study the intervals where it does not satisfy some energy condition. Based on inequalities~(\ref{EC-2}) and~(\ref{EC-3}), together with considering that the solutions that we will analyze satisfy inequality~(\ref{EC-1}), and since in general the analysis can be performed only numerically, we define the following functions to find the interval in that a black hole solution violates some of the energy conditions
\begin{equation}
g(r) =  \frac{2}{r}  \frac{d m(r)}{dr} -  \frac{d^2 m(r)}{dr^2}  \, ,
\label{DEC-C1}
\end{equation}

\begin{equation}
h(r) =  \frac{2}{r}  \frac{d m(r)}{dr} +  \frac{d^2 m(r)}{dr^2}  \, .
\label{DEC-C2}
\end{equation}

\subsection{Case I: solution that satisfies the DEC everywhere}
The first model is taken from Ref.~\cite{Balart:2014jia}, which we call case I in this article. The metric function in this model is
\begin{equation}
f_I(r) = 1 - \frac{2 M}{r}\left(1-  \frac{1}{\left( 1+ \left(\frac{2 M r}{q^2}\right)^{3}\right)^{1/3}} \right) \, .
\label{BH-1}
\end{equation}
This metric function asymptotically behaves as the Reissner-Nordström solution. In the limit $r \rightarrow 0$ behaves as a de Sitter solution,
\begin{equation}
f_I(r) \rightarrow 1 - \frac{16 M^4}{3 q^6}r^2  \, .
\label{}
\end{equation}
The extreme black hole occurs when $q_{ext} = 1.0257 M$.

In this case by construction the black hole solution satisfies the WEC in all $r$ and, moreover, also satisfies the DEC for $0 < q \leq q_{ext }$~\cite{Balart:2014jia}.

\subsection{Case II: ABG solution}
The second model is the ABG~\cite{AyonBeato:1998ub} solution, whose metric function is given by
\begin{equation}
f_{II}(r) = 1 - \frac{2 M}{r}\left(\frac{r^3}{(r^2 + q^2)^{3/2}} - \frac{q^2 r^3}{2 M (r^2 + q^2)^{2}} \right) \, .
\label{}
\end{equation}
This model also behaves as the Reissner-Nordström solution in the limit of weak fields, while for $r \rightarrow 0$ asymptotically it is
\begin{equation}
f_{II}(r) \rightarrow 1 - \frac{(2M-q)}{q^3}r^2  \, .
\label{}
\end{equation}
The extremal solution is obtained when $q_{ext } = 0.6342 M$.

\begin{figure}[h!]
\centering
\includegraphics[scale=0.65]{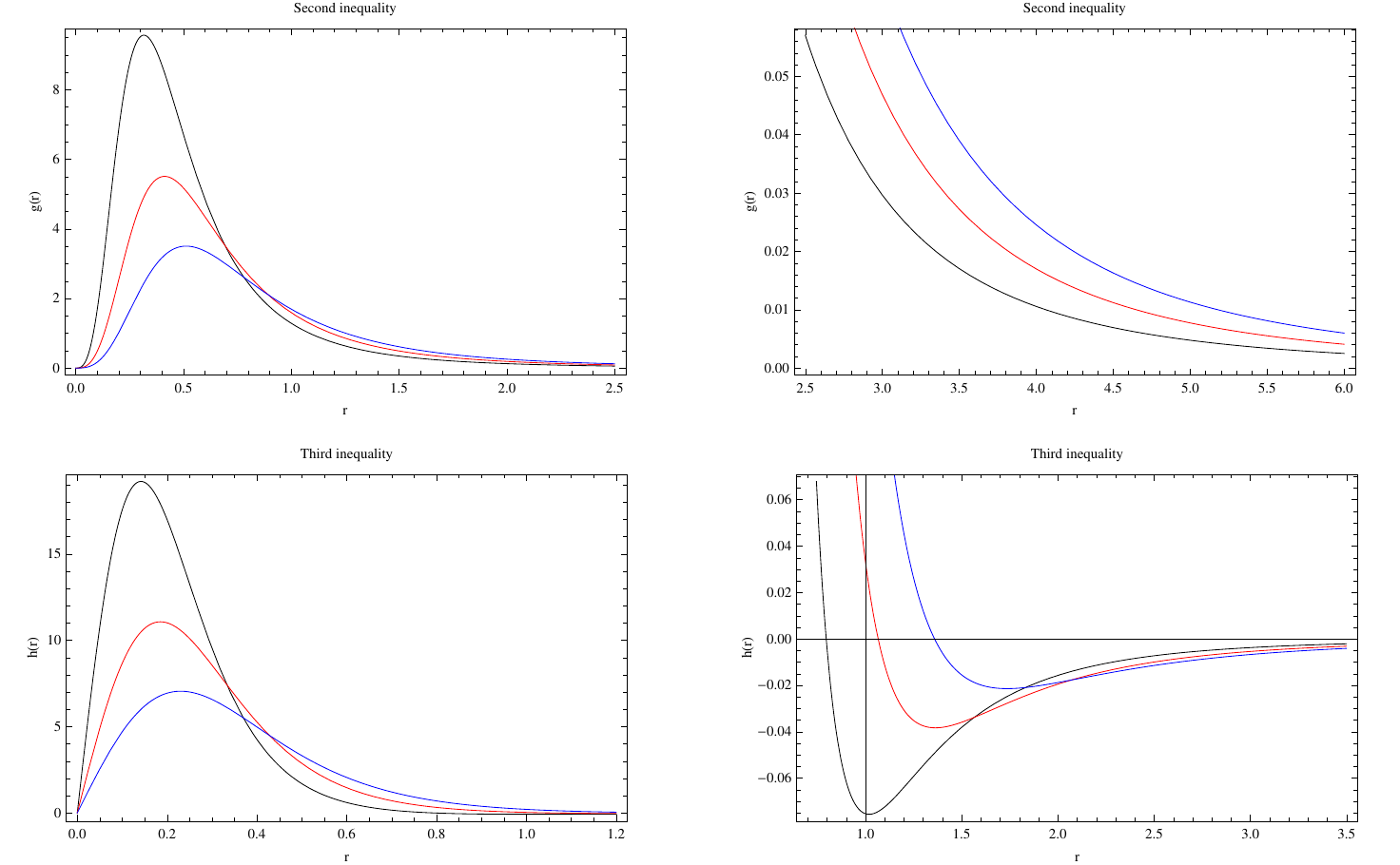}
\caption{For case II, the functions $g(r)$ and $h(r)$ are represented for $q = 0.35 M$ (black), $q = 0.45 M$ (red) and $q = 0.55 M$ (blue), and $M = 1$ in each picture. The figures on the right amplify the scale of the vertical axis. 
Here the WEC is satisfied everywhere as illustrated in the two pictures on the top, i.e. the second inequality given by Eq.~(\ref{EC-2}) is satisfied everywhere. However, inequality~(\ref{EC-3}) does not hold in an interval from $r < r_h$ to infinity, that is, the DEC is violated in this region, as can be seen in the fourth figure, where the curves cross towards down the $r$ axis.}
\label{DEC-case2}
\end{figure}

As is known, the ABG solution satisfies the WEC everywhere (i.e. inequality~(\ref{EC-1}) and condition $g(r) > 0$ are satisfied for all $r$), however from a value of $r$ to infinity it violates the DEC, since $h(r) < 0$, as shown in Fig.~(\ref{DEC-case2}).

The DEC can be violated from a point that is inside the event horizon or from a point that is outside, to infinity. This is determined by the value of the electric charge of the black hole. In this way, if the electric charge is $q_0 = 0.5727 M$ (which determines that the event horizon is at $r_{h0} = 1.4288 M$) the DEC is violated from the event horizon to infinity. If $q < q_0$ the DEC is violated from a point inside the event horizon (which is different from $r_{h0}$). And if $q_0 < q \leq q_{ext }$ the point lies outside the event horizon.
\label{Figur}

\subsection{Case III: solution with exponential}
The following model is one of the solutions proposed in Ref.~\cite{Balart:2014cga}, and whose metric function is written as
\begin{equation}
f_{III}(r) = 1 - \frac{2 M}{r}\exp\left(-\frac{q^2}{2 M r}\right)\, ,
\label{}
\end{equation}
which behaves like the Reissner-Nordström solution in the limit $r \rightarrow \infty$. The extreme case is obtained when $q_{ext } = 1.2131 M$ and $r_h =0.7355 M$.

\begin{figure}[h!]
\centering
\includegraphics[scale=0.65]{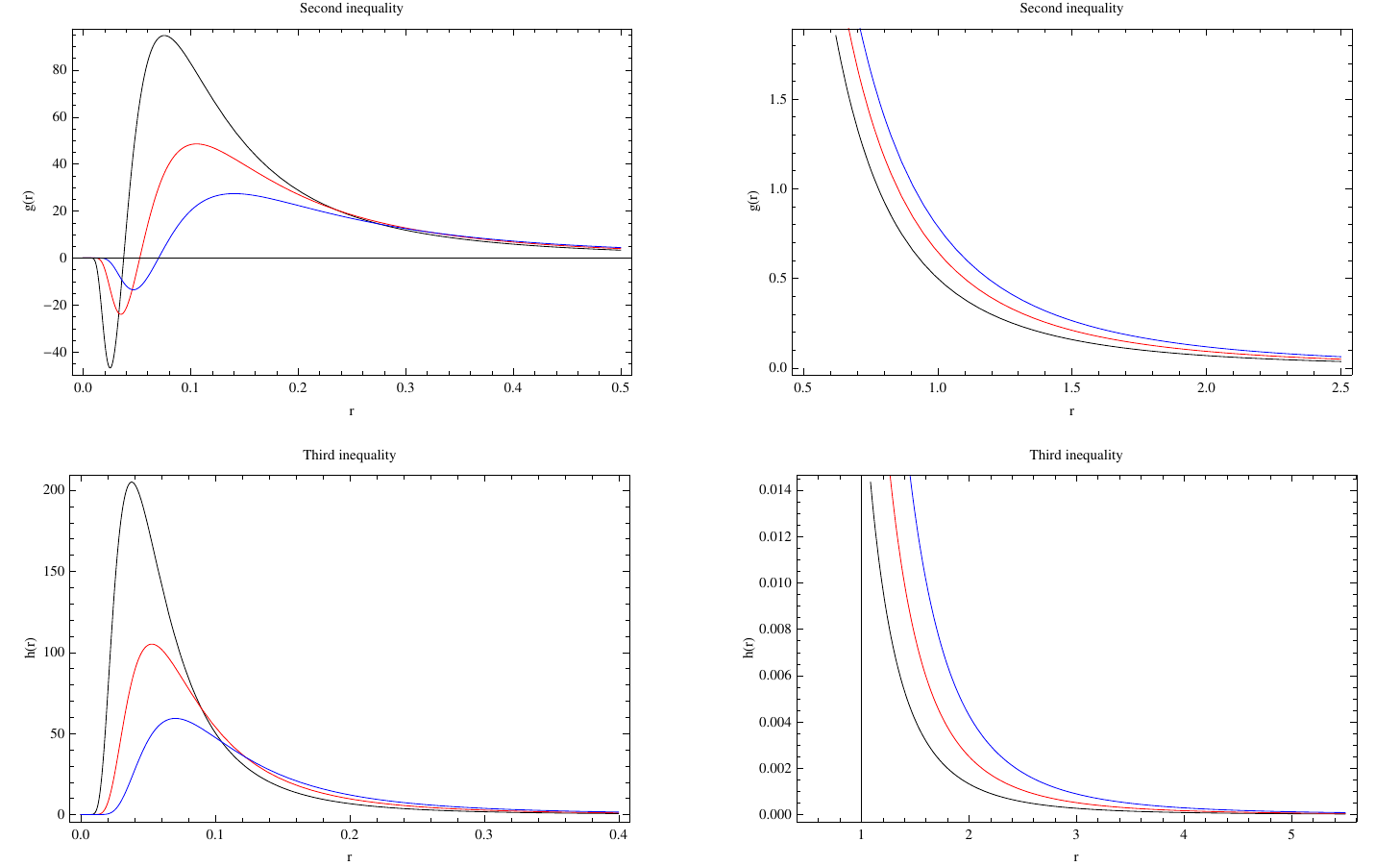}
\caption{Case III. The functions $g(r)$ and $h(r)$ are represented for $q = 0.55 M$ (black), $q = 0.65 M$ (red) and $q = 0.75 M$ (blue), and $M = $1 in each picture.
Here the condition~(\ref{EC-2}) is not satisfied in a region inside the event horizon, as seen in the first picture. The regions where the curves reached negative values correspond to the intervals in the $r$ coordinate where the WEC is violated. The condition~(\ref{EC-3}) is satisfied everywhere for all $0 < q \leq q_{ext}$.}
\label{DEC-case3}
\end{figure}

This regular black hole solution violates the WEC (or $g(r) < 0$) and therefore also violates the DEC, in a region that is inside the event horizon, as illustrated in Fig.~(\ref{DEC-case3}) for three different electric charge values $q$. However, it can be shown that this conclusion holds for all $0 < q \leq q_{ext}$. Notice also in the bottom two pictures of Fig.~(\ref{DEC-case3}) that inequality~(\ref{EC-3}) holds everywhere.
\label{Figur}

\subsection{Case IV}
The fourth model is a black hole solution that we present here
\begin{equation}
f_{IV}(r) = 1-\frac{2M}{r} \left(e^{-\frac{q^2}{2 M r}}-\frac{q^3 r^4}{2M\left(q^2+r^2\right)^3}\right)
\, .
\label{}
\end{equation}
As in the other cases, it also behaves like the Reissner-Nordström solution in the weak field limit. The extreme case is obtained when $q_{ext } = 1.1697 M$ and $r_h = 0.6153 M$.

Unlike previous cases, this regular black hole solution violates the DEC in two different regions, on the one hand because in one region it does not satisfy inequality~(\ref{EC-2}) ($g(r) > 0$) while in another distinct region does not satisfy inequality~(\ref{EC-3}) ($h(r) > 0$), as illustrated in Fig.~(\ref{DEC-case3}) for three distinct values of electric charge $q$, however the conclusion is the same for all $0 < q \leq q_{ext}$.

For $q_0 = 0.656051 M$ this black hole solution has an event horizon at $r_{h0} = 1.6973 M$ from which and to infinity the third inequality is not satisfied i.e. the solution violates the DEC. 
As in case II, if $q < q_0$ the DEC is violated from a point inside the event horizon (which is different from $r_{h0}$). If $q_0 < q \leq q_{ext }$ the DEC is violated from a point that is located outside the corresponding event horizon.
\label{Figur}
\begin{figure}[h!]
\centering
\includegraphics[scale=0.7]{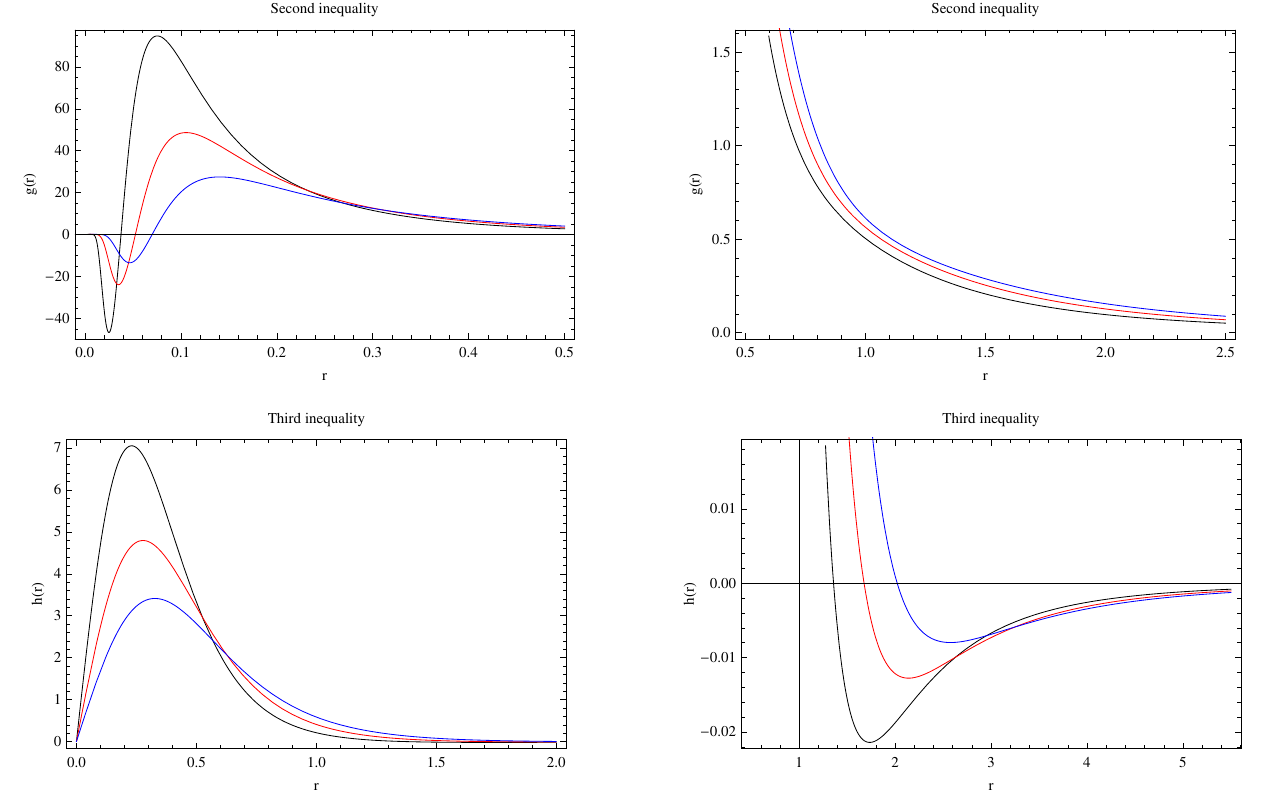}
\caption{Here are represented the functions $g(r)$ and $h(r)$ for the IV case. The first picture shows the region that is inside the event horizon, where the function $g(r)$ is negative, that is, where the WEC is violated. Additionally, in a region that is outside the event horizon and that reaches infinity, $h(r)$ is negative.
In all pictures $M = 1$, and $q = 0.55 M$ (black), $q = 0.65 M$ (red) and $q = 0.75 M$ (blue).}
\label{DEC-case4}
\end{figure}


\section{Dynamical stability}
\label{SD}

We can examine the dynamical stability of these regular black hole solutions with respect to arbitrary linear fluctuations of the metric and electromagnetic field. For this we consider the following conditions that must be satisfied by the Hamiltonian $\mathcal{H} = \mathcal{H}(x)$ for all $0 < x < x_h$~\cite{Moreno:2002gg} 
\begin{eqnarray}
\mathcal{H} &<& 0 \ ,
\label{inq-1}
\\
\mathcal{H}_x &<& 0 \ ,
\label{inq-2}
\\
\mathcal{H}_{xx} &<& 0 \ ,
\label{inq-3}
\\
3 \mathcal{H}_x &\leq& x f(x) \mathcal{H}_{xx}   \ ,
\label{inq-4}
\end{eqnarray}
here the subscript $x$ denotes differentiation with respect to the new variable $x = q^2/r^2$ and we have defined the following value $x_h =q^2/r_h^2$.

Writing the Hamiltonian of each nonlinear electrodynamics model in terms of the variable $x$ we directly verify that inequalities~(\ref{inq-1}-\ref{inq-4}) hold. In Table~\ref{tab:Ham} these Hamiltonians are listed in terms of the invariant $\mathcal{P}$ and in terms of the variable $x$, where
\begin{equation}
\mathcal{P} = \frac{1}{4} P_{\mu\nu}P^{\mu\nu} = - \frac{q^2}{2 r^4} \, .
\label{}
\end{equation}
And the following notation is used for abbreviation
\begin{equation}
s = \frac{|q|}{2 M}    \,\,\,\,\,\,\,\, , \,\,\,\,\,\,\,\, w = \left(\frac{|q|^{3/2}}{2^{3/4} M}\right)^3  \,\,\,\,\,\,\,\, \mbox{and} \,\,\,\,\,\,\,\,  U = \sqrt{-2 q^2\mathcal{P}}    \, .
\label{}
\end{equation}

\begin{table}[t]
\begin{center}
\begin{tabular}{| c | c | c | c |}
 \hline 
 Case & $\mathcal{H}(\mathcal{P})$ & $\mathcal{H}(x)$  & Reference \\  \hline
 I & $\frac{\mathcal{P}}{\left(1+w \, (-\mathcal{P})^{3/4} \right)^{4/3}}$ & $\frac{-x^2}{2 q^2\left(1+w\left(\frac{1}{2 q^2}x^2\right)^{3/4} \right)^{4/3}}$ & \cite{Balart:2014jia} \\ \hline
  II &  $\frac{\mathcal{P}(1 - 3 \,\sqrt{-2 q^2\mathcal{P}})}{1 + 2\, \sqrt{-2 q^2\mathcal{P}}} - \frac{3}{2 s q^2} \left(\frac{\sqrt{-2 q^2\mathcal{P}}}{1 + \sqrt{-2 q^2\mathcal{P}}}\right)^{5/2}$
&  $-\frac{x^2}{2 q^2}\frac{(1 - 3 x)}{(1 + x)} - \frac{3}{2 s q^2} \left(\frac{x}{1 + x}\right)^{5/2}$
 & \cite{AyonBeato:1998ub} \\ \hline   
III & $\mathcal{P} e^{-s U}$ & $-\frac{x^2}{2 q^2} e^{-s x^{1/2}}$ & \cite{Balart:2014cga} \\ \hline
IV &  $\mathcal{P} \left[e^{-s U} + \frac{2 U - 4 U^3}{\left (U^2 + 1 \right)^4} \right]$ & $-\frac{x^2}{2 q^2}\left(e^{-s x^{1/2}}+\frac{2x^{1/2}-4 x^{3/2}}{(x+1)^4}\right)$ & novel \\ \hline\end{tabular}
\caption{Hamiltonians for each source of nonlinear electrodynamics considered in this paper as functions of $\mathcal{P}$ and $x$ respectively.}
\label{tab:Ham}
\end{center}
\end{table}


\section{Quasinormal spectra}

This section is devoted to introduce essential ingredients to study scalar perturbations in regular charged black hole background. Thus, we will review the basic theory of QNMs and the numerical method to compute the corresponding frequencies.
%
%
\subsection{Wave equation for scalar perturbations}
%
Let us assume the propagation of a test real scalar field, $\Phi$, in a four dimensional gravitational background. Being $S[g_{\mu \nu} ,\Phi]$ the corresponding action, we can write the following expression 
\begin{align}
S[g_{\mu \nu} ,\Phi] \equiv \frac{1}{2} \int \mathrm{d}^4 x \sqrt{-g}
\Bigg[
\partial^{\mu} \Phi \partial_{\mu} \Phi + \xi R \Phi^2 + \frac{1}{2}m^2 \Phi^2
\Bigg]\,.
\end{align}
In what follow, we will consider an auxiliary massless  and neutral scalar field minimally coupled to gravity (i.e., we will set $\xi=0$ and $m=0$).
Now, taking advantage of the Klein-Gordon equation (see \cite{crispino,Pappas1,Pappas2,Panotopoulos:2019gtn,Rincon:2021gwd,Gonzalez:2022ote,Rincon:2020cos} and references therein)
\begin{equation}
\frac{1}{\sqrt{-g}}\partial_{\mu}\left(\sqrt{-g}g^{\mu\nu}\partial_{\nu}\Phi\right) = 0.
\end{equation}
In this simplified scenario, the corresponding Klein-Gordon equation may be solved 
following the method of separation of variables in the appropriate coordinate system. Thus, taking into consideration the symmetries of the metric tensor, we will take an ansatz for the wave function in spherical coordinates $r,\theta,\phi$ as follows
\begin{equation}
\Phi(t, r, \theta, \phi) = e^{-i\omega t}\frac{\psi(r)}{r}Y_{\ell m}(\theta, \phi),
\end{equation}
where the functions $Y_{\ell m}(\theta, \phi)$ are the conventional spherical harmonics which depend on the angular coordinates only \cite{book}, and $\omega$ is the unknown frequency to be determined after impose the concrete boundary conditions. Doing so, the relevant equation is then writen as
%
%
\begin{align}
\begin{split}
& \frac{\omega^{2}r^{2}}{f(r)} + \frac{r}{\psi(r)}\frac{d}{dr}\left[r^{2}f(r)\frac{d}{dr}\left(\frac{\psi(r)}{r}\right)\right] 
- \ell(\ell + 1) = 0.
\label{KG}
\end{split}
\end{align}
In the latter equation we have used the angular part, i.e., 
\begin{align}
    \begin{split}
&\frac{1}{\sin\theta}\frac{\partial}{\partial\theta}\left(\sin\theta\frac{\partial Y(\Omega)}{\partial\theta}\right) + \frac{1}{\sin^{2}\theta}\frac{\partial^{2}Y(\Omega)}{\partial\phi^{2}} = 
-\ell(\ell + 1)Y(\Omega),
\end{split}
\end{align}
with $\ell(\ell + 1)$ being the corresponding eigenvalue, and $\ell$ is the angular degree. 
The Klein-Gordon equation (Eq. (\ref{KG})) can be rewriten in term of the tortoise coordinate $r_{*}$ in the Schr{\"o}dinger-like form:
\begin{equation}
\frac{\mathrm{d}^{2}\psi(r)}{\mathrm{d}r_{*}^{2}} + \left[\omega^{2} - V(r)\right]\psi(r) = 0 \ ,
\end{equation}
being the standard tortoise coordinate obtained using the following relation 
\begin{align}
    r_{*}  \equiv  \int \frac{\mathrm{d}r}{f(r)} \ ,
\end{align}
while the effective potential barrier, $V(r)$, is computed to be
\begin{equation}
V(r) = f(r)
\Bigg[ 
\frac{\ell(\ell + 1)}{r^{2}} + \frac{f'(r)}{r}
\Bigg] \ .
\label{poten}
\end{equation}
and the prime denotes a derivative with respect to $r$. To complete this part, we should specify the boundary conditions.
Finally, the wave equation must be supplemented by the following boundary conditions
\begin{equation}
\Phi \rightarrow \: \exp( i \omega r_*), \; \; \; \; \; \; r_* \rightarrow - \infty \ ,
\end{equation}
\begin{equation}
\Phi \rightarrow \: \exp(-i \omega r_*), \; \; \; \; \; \; r_* \rightarrow  \infty \ .
\end{equation}
Thus, as can be observed from the time dependence of the scalar wave function, $\Phi \sim \exp(-i \omega t)$, a frequency with a negative imaginary part means a decaying (stable) mode, whereas a frequency with a positive imaginary part means an increasing (unstable) mode.

\begin{figure*}[ht!]
\centering
\includegraphics[scale=0.81]{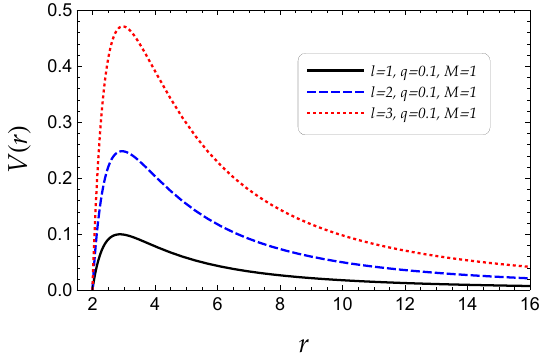} \
\includegraphics[scale=0.81]{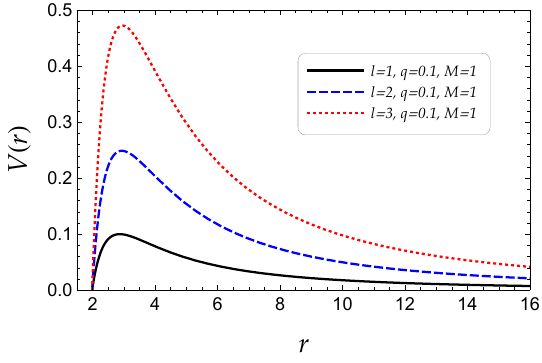} \
\\
\includegraphics[scale=0.81]{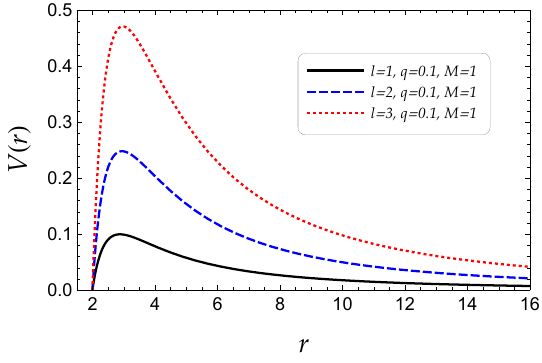} \
\includegraphics[scale=0.81]{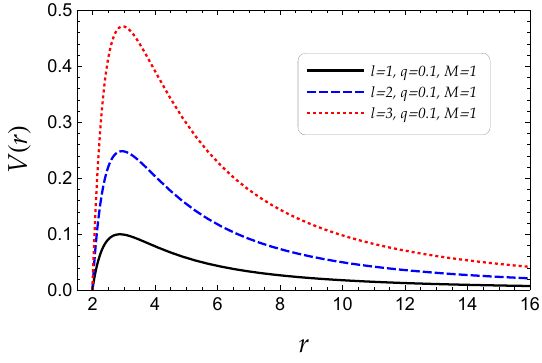} \
\caption{
Effective potential for scalar perturbations against the radial coordinate for $q=0.1$, $M=1$ and $\ell=\{1, 2, 3\}$. Top-left panel corresponds to the first model, top-right panel corresponds to the second model, bottom-left panel corresponds to the third model and bottom-right panel corresponds to the fourth model.
}
\label{fig:1} 	
\end{figure*}

In Fig.~\eqref{fig:1}, we show the effective potential for the four models discussed above. Thus, top-left panel corresponds to the first model, top-right panel corresponds to the second model, bottom-left panel corresponds to the third model and bottom-right panel corresponds to the fourth model. Also, 
The qualitative behavior of the effective potential for all cases is basically the same for the numerical values used. We have fixed the electric charge $q$ and the mass $M$ when the angular degree $\ell$ varies.
Thus, we notice that when we increase the angular number, $\ell$, the effective potential increases too. Similarly, when $\ell$ increases, $r_{\text{max}}$ is modified and slightly shifted to the right (in all cases).
%
%
%
%
%
%

\subsection{Numerical computation via WKB method}

Quasinormal modes are usually computed i) analytically (when it is possible), and ii) numerically. In particular, analytical computations are possible for instance:
i) when the effective potential barrier takes the form of the well-known P{\"o}schl-Teller potential \cite{potential,ferrari,cardoso2,lemos,molina,panotop1}, 
or 
ii) when the differential equation (for the radial part of the wave function) can be recast into the Gauss' hypergeometric function \cite{exact1,exact2,exact3,Gonzalez:2010vv,exact4,exact5,exact6}. 

It is very well-known that, in general, an exact analytic solution is not possible to achieve. The reason of that is the complexity and non-trivial structure of the differential equation involved. By the above reasons, it is usually necessary to employ one of the numerical approaches available in the literature. 
There are a large number of methods used to obtain, in an accuracy, the QN spectra of black holes. We can mention, for example: 
  i) the Frobenius method \cite{Horowitz:1999jd,Starinets:2002br}, 
 ii) the generalization of the Frobenius series, 
iii) fit and interpolation approach, 
 iv) the method of continued fraction \cite{Leaver:1985ax,leJMP,Leaver:1986gd}, 
  v) the asymptotic iteration method \cite{Cho:2011sf,2003JPhA...3611807C,Ciftci:2005xn}
although more methods are known to exist. 
For more details the interested reader may consult for instance \cite{review3}.

\smallskip

Thus, although there are many numerical methods to obtain the QNMs, we shall focus here on one of the most popular
up to now, i.e., the WKB semi-classical method (commonly implemented in elementary 1D quantum 
mechanical problems). The formalism and the details have been investigated extensively by many authors, 
increasing the order of the approximation. 
In light of the WKB semi-classical approximation is well-known \cite{wkb0,wkb1,wkb2,wkb3,wkb4}, 
we will circumvent the inclusion of details.  
Following the WKB method, the QN spectra may be computed via the following expression
\begin{equation}
\omega_n^2 = V_0+(-2V_0'')^{1/2} \Lambda(n) - i \nu (-2V_0'')^{1/2} [1+\Omega(n)]\,,
\end{equation}
where 
i) $V_0''$ represent the second derivative of the potential (at the maximum), 
ii) $\nu=n+1/2$, $V_0$ is the maximum of the effective potential barrier, 
iii) $n=0,1,2...$ symbolizes the overtone number, 
while the functions
$\Lambda(n), \Omega(n)$ are quite long and intricate expressions of $\nu$ (and derivatives of the potential at the maximum). The concrete form of above symbols can be found in \cite{wkb3}. 

\smallskip

Notice that the 3rd order approximation was first constructed by Iyer and Will in \cite{wkb1}, and after that extended to higher orders. Thus, to perform our computations, we have used here a Wolfram Mathematica \cite{wolfram} notebook utilizing the WKB method at any order from one to six \cite{code}. 

\smallskip

In the present work we have implemented the WKB method at sixth order. Besides, it should be mentioned that for a given angular degree, $\ell$, we have considered values $n < \ell$ only, since it is known that the method works well for high angular degrees. Higher order in the WKB approximation have been investigated in \cite{Opala,Konoplya:2019hlu,RefExtra2} where a recipe for simple, quick, efficient and accurate computations was provided. Moreover, notice that as the WKB series converges only asymptotically, there is no mathematically strict criterion for evaluation of an error according to \cite{Konoplya:2019hlu}. However, the sixth/seventh order usually produces the best results. In a future study we shall perform an error analysis and employ the Pad{\'e} approximants with the hope to achieve a higher accuracy.

\smallskip

\begin{figure*}[ht!]
\centering
\includegraphics[scale=0.81]{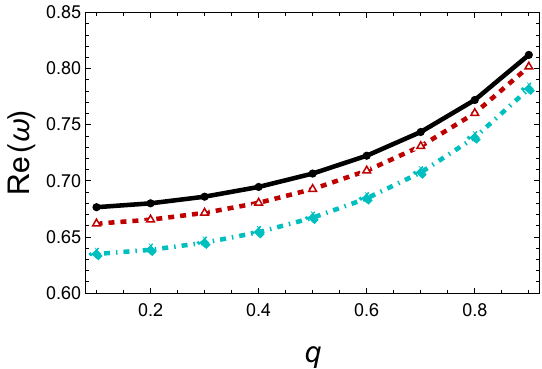} \
\includegraphics[scale=0.81]{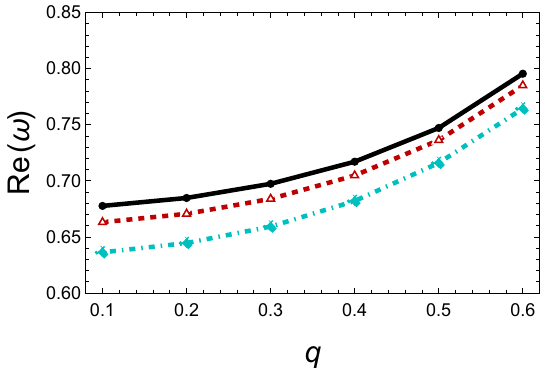}

\vskip 0.75cm 

\includegraphics[scale=0.81]{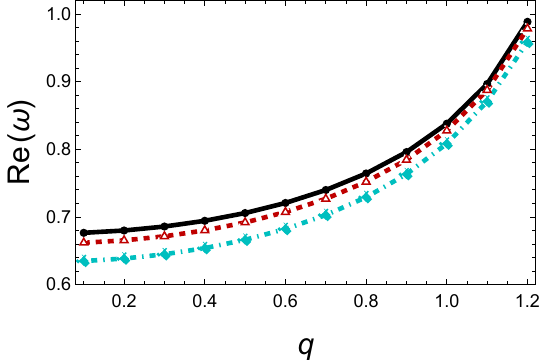} \
\includegraphics[scale=0.81]{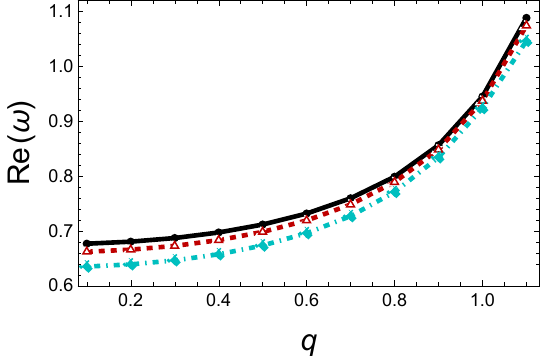}
\caption{
QNMs for all cases investigated with $M=1$, $\ell=3$ and $n=\{ 0, 1, 2 \}$. 
Real part of $\omega$ against the electric charge $q$ for case I to IV from top left to bottom right.
The color code is:
  i) solid black line for $n=0$,
 ii) dashed red line for $n=1$ and
iii) dot-dashed cyan line for $n=2$.
}
\label{fig:modes}	
\end{figure*}

\begin{figure*}[ht!]
\centering
\includegraphics[scale=0.81]{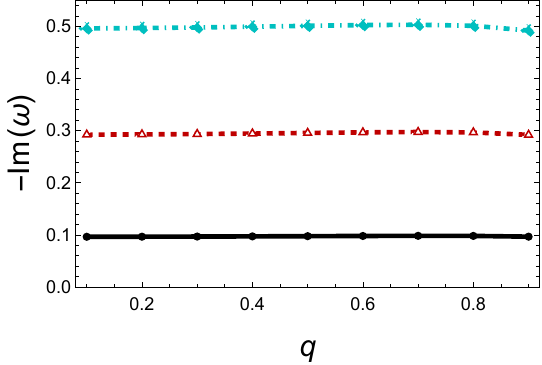} \
\includegraphics[scale=0.81]{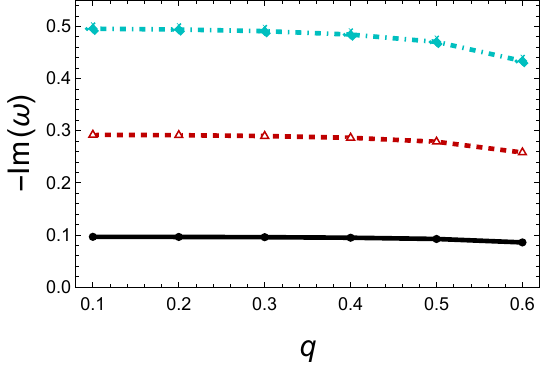} 

\vskip0.75cm

\includegraphics[scale=0.81]{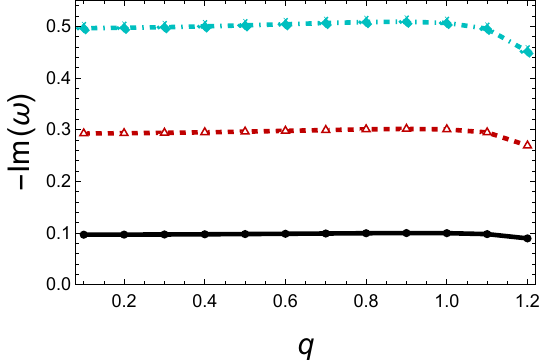} \
\includegraphics[scale=0.81]{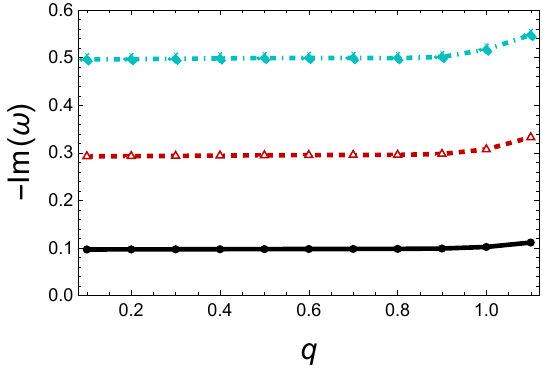} 
\caption{
QNMs for all cases investigated with $M=1$, $\ell=3$ and $n=\{ 0, 1, 2 \}$. 
Imaginary part of $\omega$ against the electric charge $q$ for case I to IV from top left to bottom right.
The color code is:
  i) solid black line for $n=0$,
 ii) dashed red line for $n=1$ and
iii) dot-dashed cyan line for $n=2$.
}
\label{fig:modes1}	
\end{figure*}

\begin{figure*}[ht!]
\centering
\includegraphics[scale=0.81]{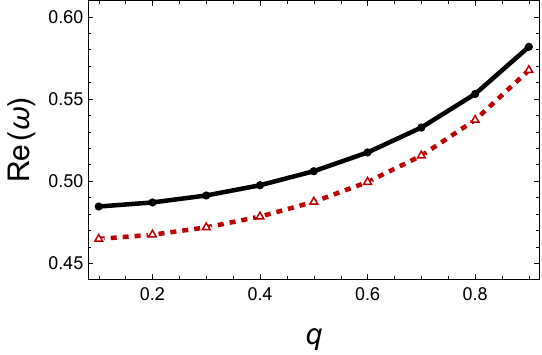} \
\includegraphics[scale=0.81]{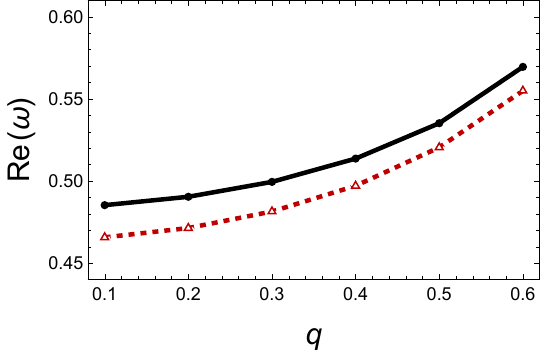} 

\vskip0.75cm

\includegraphics[scale=0.81]{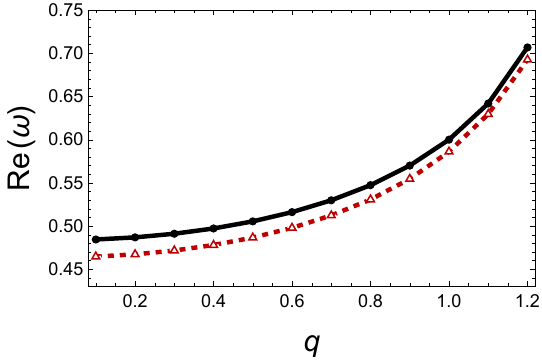} \
\includegraphics[scale=0.81]{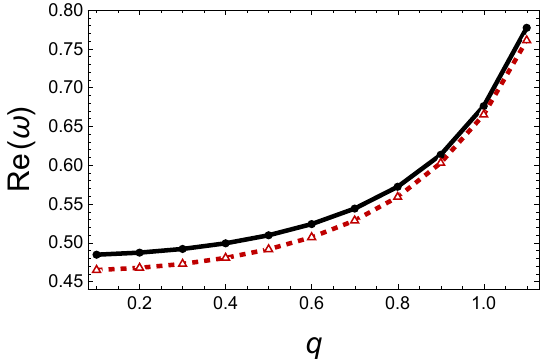} 
\caption{
QNMs for all cases investigated with $M=1$, $\ell=2$ and $n=\{ 0, 1 \}$. 
Real part of $\omega$ against the electric charge $q$ for case I to IV from top left to bottom right.
The color code is:
  i) solid black line for $n=0$ and
 ii) dashed red line for $n=1$. 
}
\label{fig:modes2}	
\end{figure*}

\begin{figure*}[ht!]
\centering
\includegraphics[scale=0.81]{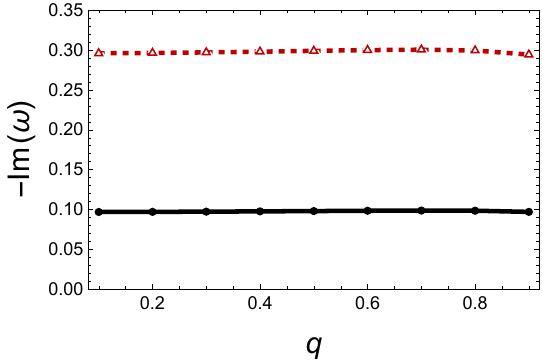} \
\includegraphics[scale=0.81]{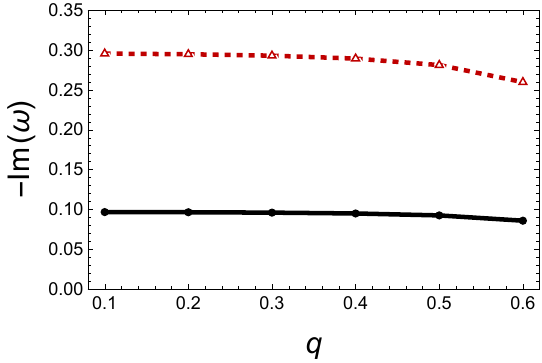} 

\vskip0.75cm

\includegraphics[scale=0.81]{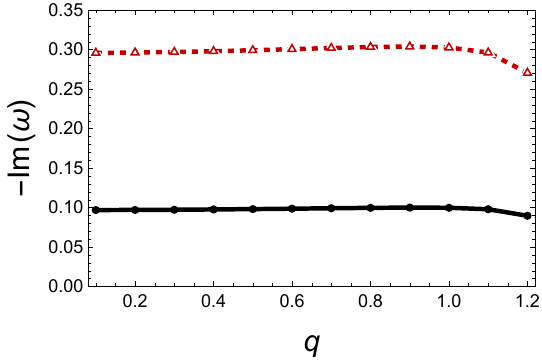} \
\includegraphics[scale=0.81]{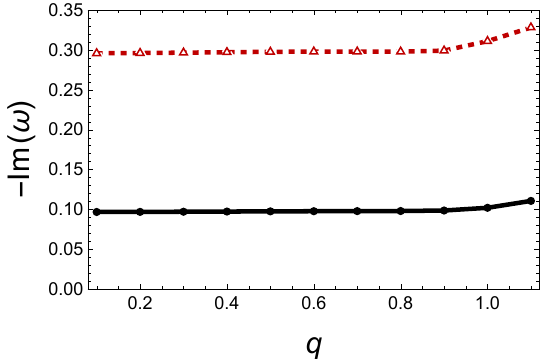} 
\caption{
QNMs for all cases investigated with $M=1$, $\ell=2$ and $n=\{ 0, 1 \}$. 
Imaginary part of $\omega$ against the electric charge $q$ for case I to IV from top left to bottom right.
The color code is:
  i) solid black line for $n=0$ and
 ii) dashed red line for $n=1$. 
}
\label{fig:modes2b}	
\end{figure*}

\begin{figure*}[ht!]
\centering
\includegraphics[scale=0.81]{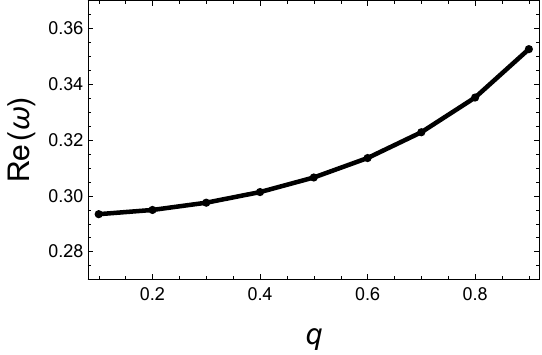} \
\includegraphics[scale=0.81]{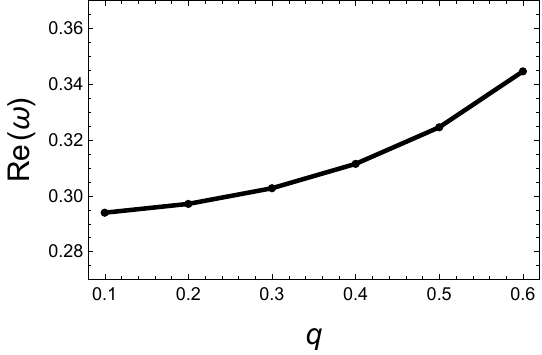} 

\vskip0.75cm

\includegraphics[scale=0.81]{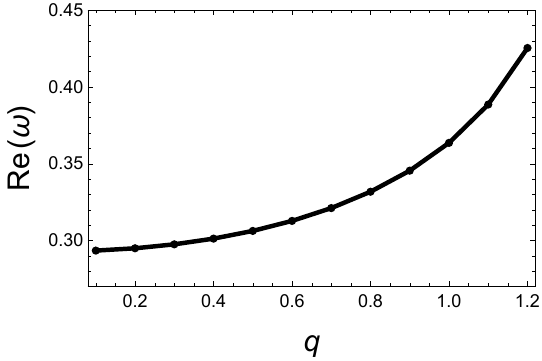} \
\includegraphics[scale=0.81]{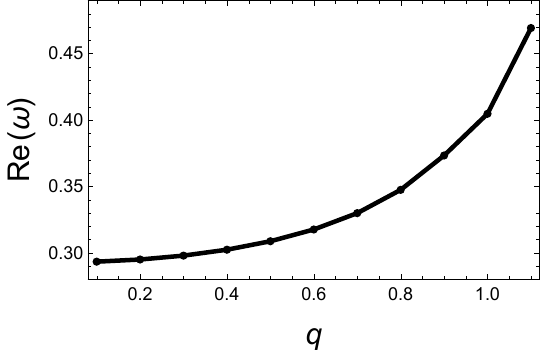} 
\caption{
QNMs for all cases investigated with $M=1$, $\ell=1$ and $n=0$. 
Real part of $\omega$ against the electric charge $q$ for case I to IV from top left to bottom right.
}
\label{fig:modes3}	
\end{figure*}

\begin{figure*}[ht!]
\centering
\includegraphics[scale=0.81]{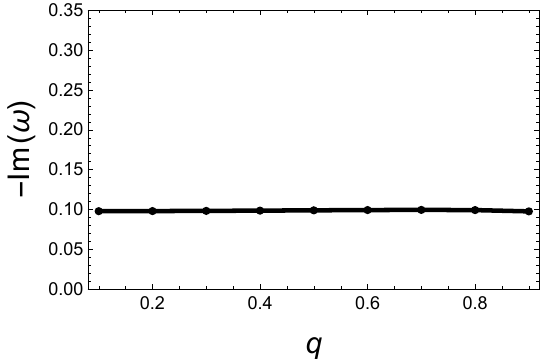} \
\includegraphics[scale=0.81]{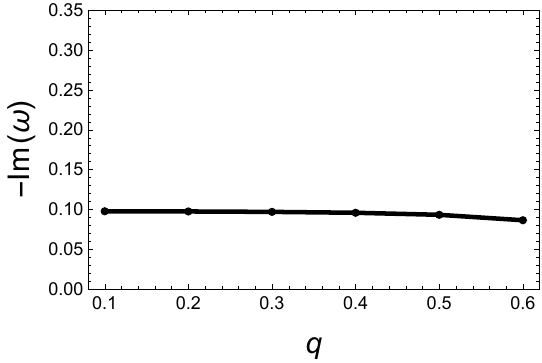} 

\vskip0.75cm

\includegraphics[scale=0.81]{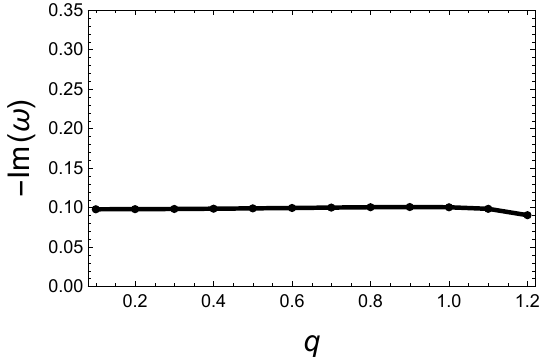} \
\includegraphics[scale=0.81]{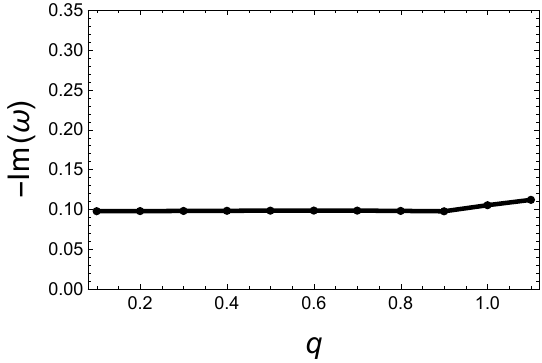} 
\caption{
QNMs for all cases investigated with $M=1$, $\ell=1$ and $n=0$. 
Imaginary part of $\omega$ against the electric charge $q$ for case I to IV from top left to bottom right.
}
\label{fig:modes3b}	
\end{figure*}

Our main results, to be discussed below, are summarized in figures~(\ref{fig:modes})-(\ref{fig:eikonal}) and tables (\ref{table:First set})-(\ref{table:Fourth set}). The QNMs for scalar perturbations of the model III was studied in \cite{Panotopoulos:2019qjk}, although in that work energy conditions were not discussed. In the present work, we show the spectra for all 4 models for comparison reasons. Finally, for recent studies of QNMs in alternative backgrounds see \cite{Okyay:2021nnh,Ovgun:2018gwt,Pantig:2022gih,Gonzalez:2021vwp,Panotopoulos:2020mii,Rincon:2020iwy,Rincon:2018sgd,Panotopoulos:2017hns,Avalos:2023jeh} and references therein.

%


%
%


\begin{table}[ph!]
\centering
\caption{Quasinormal frequencies (varying $\ell$, $n$ and $q$) with $M=1$ for the first model considered in this work. 
}
{
\begin{tabular}{c|c|ccc} 
\toprule
$q$ & $n$ &  $\ell=1$ & $\ell=2$ & $\ell=3$ 
\\ \toprule
\hline
    & 0 &  0.293407 - 0.0978114 i  & 0.484455 - 0.0968185 i & 0.676499 - 0.0965534 i \\
0.1 & 1 &      			  		   & 0.464698 - 0.2957730 i & 0.661832 - 0.2924410 i  \\
    & 2 &  				  		   &  			            & 0.634804 - 0.4962460 i
\\ \toprule
\hline
    & 0 &   0.294919 - 0.0979588 i  & 0.486929 - 0.0969738 i & 0.679946 - 0.0967102 i \\
0.2 & 1 &   				   		& 0.467290 - 0.2962050 i & 0.665366 - 0.2928930 i  \\
    & 2 &    				   		&  			             & 0.638500 - 0.4969420 i
\\ \toprule
\hline
    & 0 &   0.297517 - 0.0981972 i  & 0.491179 - 0.0972258 i & 0.685866 - 0.0969647 i \\
0.3 & 1 &  				  		    & 0.471746 - 0.2969000 i & 0.671440 - 0.2936260 i  \\
    & 2 &  				   			&  			     		 & 0.644858 - 0.4980600 i
\\ \toprule
\hline
    & 0 &  0.301325 - 0.0985124 i    & 0.497411 - 0.0975606 i  & 0.694549 - 0.0973033 i \\
0.4 & 1 &  	  				  	  	 & 0.478295 - 0.2978110 i  & 0.680357 - 0.2945940 i  \\
    & 2 &  			   			  	 &  			     	   & 0.654209 - 0.4995150 i
\\ \toprule
\hline
    & 0 &   0.306550 - 0.098875 i    & 0.505966 - 0.0979498 i & 0.706468 - 0.0976983 i \\
0.5 & 1 &   				  	    & 0.487305 - 0.2988430 i & 0.692614 - 0.2957090 i  \\
    & 2 &  	 				   		&  			     		 & 0.667090 - 0.5011440 i
\\ \toprule
\hline
    & 0 &   0.313519 - 0.0992245 i  & 0.517383 - 0.098335 i  & 0.722378 - 0.098092 i \\
0.6 & 1 &   				   		& 0.499367 - 0.299802 i  & 0.709003 - 0.296789 i  \\
    & 2 &  	 				   		&  			     		 & 0.684359 - 0.502612 i
\\ \toprule
\hline
    & 0 &   0.322760 - 0.099430 i   & 0.532545 - 0.0985892 i  & 0.743515 - 0.0983592 i \\
0.7 & 1 &   				   		& 0.515433 - 0.3002770 i  & 0.730816 - 0.2974410 i  \\
    & 2 &  	 				   		&  			     		  & 0.707399 - 0.5032100 i
\\ \toprule
\hline
    & 0 &   0.335193 - 0.0991747 i     & 0.55299 - 0.0984097 i  & 0.772039 - 0.0982003 i \\
0.8 & 1 &   				   		& 0.537143 - 0.299286 i & 0.760287 - 0.296729 i  \\
    & 2 &  	 				   		&  			     		  & 0.738566 - 0.501265 i
\\ \toprule
\hline
    & 0 &   0.352528 - 0.0975726 i     & 0.581718 - 0.0969387 i  & 0.812206 - 0.0967641 i \\
0.9 & 1 &   				   		& 0.567456 - 0.294097 i  & 0.801669 - 0.292015 i  \\
    & 2 &  	 				   		&  			     		  & 0.781991 - 0.492131 i
\\ \toprule
\hline
\end{tabular} 
\label{table:First set}
}
\end{table}

\begin{table}[ph!]
\centering
\caption{Quasinormal frequencies (varying $\ell$, $n$ and $q$) with $M=1$ for the second model considered in this work. 
}
{
\begin{tabular}{c|c|ccc} 
\toprule
$q$ & $n$ &  $\ell=1$ & $\ell=2$ & $\ell=3$ 
\\ \toprule
\hline
    & 0 &  0.293931 - 0.0976968 i  & 0.485289 - 0.0967053 i & 0.677650 - 0.0964418 i \\
0.1 & 1 &      			  		   & 0.465678 - 0.2953930 i & 0.663092 - 0.2920850 i  \\
    & 2 &  				  		   &  			            & 0.636265 - 0.4955830 i
\\ \toprule
\hline
    & 0 &   0.297089 - 0.0974668 i  & 0.490386 - 0.0964901 i & 0.684725 - 0.096233 i \\
0.2 & 1 &   				   		& 0.471346 - 0.2945840 i & 0.670586 - 0.291373 i  \\
    & 2 &    				   		&  			             & 0.644536 - 0.494117 i
\\ \toprule
\hline
    & 0 &   0.302702 - 0.0969435 i  & 0.499465 - 0.0960008 i & 0.697331 - 0.0957558 i \\
0.3 & 1 &  				  		    & 0.481419 - 0.2928130 i & 0.683930 - 0.2897840 i  \\
    & 2 &  				   			&  			     		 & 0.659232 - 0.4909520 i
\\ \toprule
\hline
    & 0 &  0.311423 - 0.095815 i  & 0.513636 - 0.0949421 i & 0.717035 - 0.0947178 i \\
0.4 & 1 &  	  				  	  & 0.497062 - 0.2891280 i & 0.704735 - 0.2864040 i  \\
    & 2 &  			   			  &  			     	   & 0.682021 - 0.4844610 i
\\ \toprule
\hline
    & 0 &   0.324537 - 0.0932644 i  & 0.535188 - 0.0925256 i & 0.747094 - 0.0923352 i \\
0.5 & 1 &   				  	    & 0.520478 - 0.2810110 i & 0.736221 - 0.2788030 i  \\
    & 2 &  	 				   		&  			     		 & 0.715954 - 0.4703420 i
\\ \toprule
\hline
    & 0 &   0.344567 - 0.0864543 i  & 0.569504 - 0.0858761 i & 0.795453 - 0.0857209 i \\
0.6 & 1 &   				   		& 0.554955 - 0.2596170 i & 0.784936 - 0.2581850 i  \\
    & 2 &  	 				   		&  			     		 & 0.764387 - 0.4336200 i
\\ \toprule
\hline
\end{tabular} 
\label{table:Second set}
}
\end{table}

\begin{table}[ph!]
\centering
\caption{Quasinormal frequencies (varying $\ell$, $n$ and $q$) with $M=1$ for the third model considered in this work. 
}
{
\begin{tabular}{c|c|ccc} 
\toprule
$q$ & $n$ &  $\ell=1$ & $\ell=2$ & $\ell=3$ 
\\ \toprule
\hline
    & 0 &  0.293406 - 0.0978115 i  & 0.484454 - 0.0968186 i & 0.676498 - 0.0965535 i \\
0.1 & 1 &      			  		   & 0.464697 - 0.2957730 i & 0.661831 - 0.2924410 i  \\
    & 2 &  				  		   &  			            & 0.634803 - 0.4962470 i
\\ \toprule
\hline
    & 0 &   0.294912 - 0.0979604 i  & 0.486918 - 0.0969754 i & 0.679930 - 0.0967117 i \\
0.2 & 1 &   				   		& 0.467276 - 0.2962100 i & 0.665349 - 0.2928980 i  \\
    & 2 &    				   		&  			             & 0.638480 - 0.4969520 i
\\ \toprule
\hline
    & 0 &   0.297479 - 0.0982059 i  & 0.491119 - 0.0972344 i & 0.685783 - 0.0969731 i \\
0.3 & 1 &  				  		    & 0.471676 - 0.2969280 i & 0.671349 - 0.2936530 i  \\
    & 2 &  				   			&  			     		 & 0.644753 - 0.4981100 i
\\ \toprule
\hline
    & 0 &  0.301197 - 0.098543 i     & 0.497208 - 0.0975909 i  & 0.694268 - 0.0973332 i \\
0.4 & 1 &  	  				  	  	 & 0.478056 - 0.2979120 i  & 0.680049 - 0.2946890 i  \\
    & 2 &  			   			  	 &  			     	   & 0.653853 - 0.4996910 i
\\ \toprule
\hline
    & 0 &   0.306211 - 0.0989618 i  & 0.505422 - 0.0980355 i & 0.705716 - 0.0977828 i \\
0.5 & 1 &   				  	    & 0.486670 - 0.2991280 i & 0.691794 - 0.2959770 i  \\
    & 2 &  	 				   		&  			     		 & 0.666146 - 0.5016380 i
\\ \toprule
\hline
    & 0 &   0.312731 - 0.0994435 i  & 0.516118 - 0.0985504 i  & 0.720626 - 0.0983045 i \\
0.6 & 1 &   				   		& 0.497895 - 0.3005140 i  & 0.707097 - 0.2974610 i  \\
    & 2 &  	 				   		&  			     		  & 0.682173 - 0.5038440 i
\\ \toprule
\hline
    & 0 &   0.321075 - 0.0999512 i  & 0.529823 - 0.0991003 i  & 0.739737 - 0.0988635 i \\
0.7 & 1 &   				   		& 0.512292 - 0.3019550 i  & 0.726725 - 0.2990300 i  \\
    & 2 &  	 				   		&  			     		  & 0.702747 - 0.5061070 i
\\ \toprule
\hline
    & 0 &   0.331724 - 0.100411 i   & 0.547347 - 0.099614 i  & 0.764186 - 0.0993894 i \\
0.8 & 1 &   				   		& 0.530719 - 0.303217 i  & 0.751847 - 0.3004640 i  \\
    & 2 &  	 				   		&  			     		 & 0.729098 - 0.5080280 i
\\ \toprule
\hline
 	& 0 &   0.345441 - 0.100667 i  & 0.569988 - 0.099938 i  & 0.795797 - 0.0997299 i \\
0.9 & 1 &   				   		& 0.554535 - 0.303809 i  & 0.784339 - 0.3012870 i  \\
    & 2 &  	 				   		&  			     		 & 0.763183 - 0.5087640 i
\\ \toprule
\hline
	& 0 &   0.363533 - 0.100345 i   & 0.599997 - 0.099705 i  & 0.837749 - 0.0995193 i \\
1.0 & 1 &   				   		& 0.586058 - 0.302566 i  & 0.827435 - 0.3003710 i  \\
    & 2 &  	 				   		&  			     		 & 0.808316 - 0.5063320 i
\\ \toprule
\hline
	& 0 &   0.388499 - 0.0983738 i  & 0.641838 - 0.0978466 i  & 0.896398 - 0.0976899 i \\
1.1 & 1 &   				   		& 0.629621 - 0.2961440 i  & 0.887419 - 0.2944380 i  \\
    & 2 &  	 				   		&  			       		  & 0.870552 - 0.4950440 i
\\ \toprule
\hline
	& 0 &   0.425303 - 0.0903157 i  & 0.706747 - 0.0895937 i  & 0.988322 - 0.0894092 i \\
1.2 & 1 &   				   		& 0.692614 - 0.2704350 i  & 0.978313 - 0.2690350 i \\
    & 2 &  	 				   		&  			       		  & 0.958506 - 0.4510810 i
\\ \toprule
\hline
\end{tabular} 
\label{table:Third set}
}
\end{table}


\begin{table}[ph!]
\centering
\caption{Quasinormal frequencies (varying $\ell$, $n$ and $q$) with $M=1$ for the fourth model considered in this work.
}
{
\begin{tabular}{c|c|ccc} 
\toprule
$q$ & $n$ &  $\ell=1$ & $\ell=2$ & $\ell=3$ 
\\ \toprule
\hline
    & 0 &  0.293424 - 0.0978077 i  & 0.484482 - 0.0968149 i & 0.676536 - 0.0965499 i \\
0.1 & 1 &      			  		   & 0.464729 - 0.2957610 i & 0.661872 - 0.2924290 i  \\
    & 2 &  				  		   &  			            & 0.634851 - 0.4962250 i
\\ \toprule
\hline
    & 0 &   0.295054 - 0.0979306 i  & 0.487144 - 0.0969461 i & 0.680242 - 0.0966827 i \\
0.2 & 1 &   				   		& 0.467541 - 0.2961110 i & 0.665689 - 0.2928060 i  \\
    & 2 &    				   		&  			             & 0.638875 - 0.4967790 i
\\ \toprule
\hline
    & 0 &   0.297972 - 0.0981052 i  & 0.491906 - 0.0971357 i & 0.686872 - 0.0968754 i \\
0.3 & 1 &  				  		    & 0.472599 - 0.2965980 i & 0.672538 - 0.2933410 i  \\
    & 2 &  				   			&  			     		 & 0.646128 - 0.4975310 i
\\ \toprule
\hline
    & 0 &  0.302427 - 0.0983017 i    & 0.499174 - 0.0973574 i  & 0.696988 - 0.0971015 i \\
0.4 & 1 &  	  				  	  	 & 0.480351 - 0.2971310 i  & 0.683013 - 0.2939520 i  \\
    & 2 &  			   			  	 &  			     	   & 0.657273 - 0.4983250 i
\\ \toprule
\hline
    & 0 &   0.308789 - 0.0984803 i  & 0.509549 - 0.0975801 i & 0.711427 - 0.0973296 i \\
0.5 & 1 &   				  	    & 0.491464 - 0.2976050 i & 0.698001 - 0.2945390 i  \\
    & 2 &  	 				   		&  			     		 & 0.673285 - 0.4989810 i
\\ \toprule
\hline
    & 0 &   0.317636 - 0.0985835 i  & 0.523963 - 0.0977681 i  & 0.731492 - 0.0975242 i \\
0.6 & 1 &   				   		& 0.506959 - 0.2979010 i & 0.718870 - 0.2949910 i  \\
    & 2 &  	 				   		&  			     		  & 0.695657 - 0.4993020 i
\\ \toprule
\hline
    & 0 &   0.329935 - 0.0985228 i  & 0.543939 - 0.0978954 i  & 0.759306 - 0.0976612 i \\
0.7 & 1 &   				   		& 0.528496 - 0.2979180 i  & 0.747840 - 0.2952340 i  \\
    & 2 &  	 				   		&  			     		  & 0.726803 - 0.4991580 i
\\ \toprule
\hline
    & 0 &   0.347447 - 0.0981599 i  & 0.572111 - 0.0980226 i  & 0.798550 - 0.0978110 i \\
0.8 & 1 &   				   		& 0.558897 - 0.2977960 i  & 0.788711 - 0.2954830 i  \\
    & 2 &  	 				   		&  			     		  & 0.770751 - 0.4989200 i
\\ \toprule
\hline
    & 0 &  0.373304 - 0.0976706 i   & 0.613285 - 0.0986527 i  & 0.856032 - 0.0985011 i \\
0.9 & 1 &   				   		& 0.602778 - 0.2991690 i  & 0.848146 - 0.2973280 i  \\
    & 2 &  	 				   		&  			     		  & 0.833914 - 0.5013860 i
\\ \toprule
\hline
    & 0 &   0.404708 - 0.1053030 i   & 0.676058 - 0.1020630 i  & 0.944645 - 0.1018570 i \\
1.0 & 1 &   				   		& 0.665054 - 0.3110080 i  & 0.936953 - 0.3070170 i  \\
    & 2 &  	 				   		&  			     		 & 0.923853 - 0.5169260 i
\\ \toprule
\hline
    & 0 &   0.46919 - 0.1120120 i   & 0.777111 - 0.1107050 i  & 1.088030 - 0.1111060 i \\
1.1 & 1 &   				   		& 0.760849 - 0.3281150 i  & 1.074010 - 0.3323430 i  \\
    & 2 &  	 				   		&  			       		 & 1.046230 - 0.5477710 i
\\ \toprule
\hline
\end{tabular} 
\label{table:Fourth set}
}
\end{table}

\subsection{QNMs in the eikonal limit}

\begin{figure*}[ht!]
\centering
\includegraphics[scale=0.81]{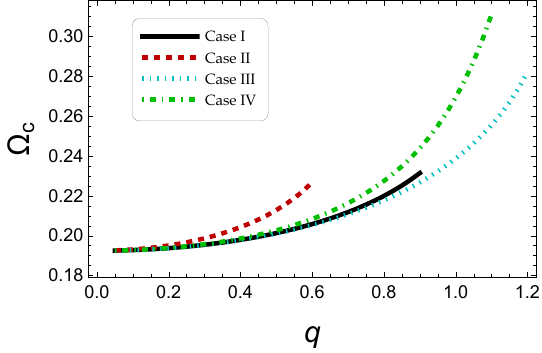} \
\includegraphics[scale=0.81]{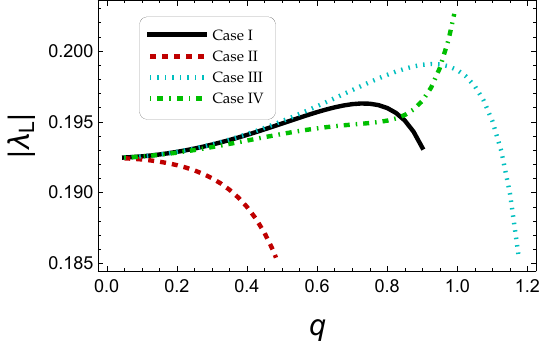} \
\caption{
QNMs in the eikonal limit:
{\bf{Left panel:}} Angular velocity vs the electric charge for $M=1$.
{\bf{Right panel:}} Lyapunov exponent againt the electric charge for $M=1$.
Both panels represent all the models, from case I to IV. 
}
\label{fig:eikonal} 	
\end{figure*}

In the eikonal regime (i.e., $\ell \gg 1$) the WKB approximation becomes increasingly accurate. Thus, we can obtain analytical expressions for the quasinormal frequencies. In such a limit ($ \ell \rightarrow \infty$), the angular momentum term that dominates in the expression for the effective potential
\begin{equation}
V(r) \approx \frac{f(r) l^2}{r^2} \equiv l^2 \sigma(r) ,
\end{equation}
where for simplicity we have defined a new function $\sigma(r) \equiv f(r)/r^2$. 
It is possible to obtain the maximum of the potential, $r_1$ by solving the following algebraic equation
\begin{equation}
2 f(r_1) - r_1 f'(r)|_{r_1} = 0.
\end{equation}
The idea and formalism was treated in \cite{eikonal1}. The QNMs, in the eikonal regime, are found to be
\begin{equation}
\omega(\ell \gg 1) = \Omega_c \ell - i \left(n+\frac{1}{2}\right) |\lambda_L| ,
\end{equation}
where the Lyapunov exponent $\lambda_L$ is given by \cite{eikonal1}
\begin{equation}
\lambda_L = r_1^2 \sqrt{\frac{\sigma''(r_1) \sigma(r_1)}{2}},
\end{equation}
while the angular velocity $\Omega_c$ at the unstable null geodesic is given by \cite{eikonal1}
\begin{equation}
\Omega_c = \frac{\sqrt{f(r_1)}}{r_1}.
\end{equation}
By applying the WKB approximation of 1st order, we can recover the same expression mentioned above for 
$\{ \Omega_c, \lambda_L \}$, see for instance \cite{Ponglertsakul:2018smo}. Thus, although in the presence of non-linear electromagnetic sources photons follow the null trajectories of particular effective geometry rather than the null geodesics of the true geometry \cite{eikonal2,eikonal3,ref2}, the previous formulas for $\Omega_c$ and for $\lambda_L$ are the same.
In particular, we know that 
i) the angular velocity determines the real part of the modes (where only the degree of angular momentum $\ell$ enters),
and
ii) the Lyapunov exponent determines the imaginary part of the modes (where only the overtone number $n$ appears).
Thus, an analytic expression for the spectrum is found where
\begin{eqnarray}
\omega_{R}(\ell \gg 1) &\equiv& \mbox{Re}(\omega) = \Omega_c \ell ,
\\
\omega_{I}(\ell \gg 1) &\equiv& \mbox{Im}(\omega) = - \left(n+\frac{1}{2}\right) |\lambda_L| ,
\end{eqnarray}
In fig. (\ref{fig:eikonal}) we present the angular velocity (left panel) and the absolute value of the Lyapunov exponent (right panel) versus $q$ assuming $M=1$ for all four models. 

As far as the real part of the frequencies is concerned, the angular velocity, $\Omega_c$, increases monotonically with the electric charge in all cases.
Regarding the imaginary part of the spectra, the Lyapunov exponent (its absolute value) exhibits a different behaviour from one model to another. In particular, in the cases of the models I and III, it reaches a maximum value first and subsequently it decreases (as opposed to the Bardeen black hole studied in \cite{correa}, where it was found that $|\lambda_L|$ decreases monotonically with $q$).
This is precisely the behaviour observed in the case of the standard RN geometry. In the case of model II, however, $|\lambda_L|$ monotonically decreases with $q$, similarly to \cite{correa}. Finally, model IV is the only case where the Lyapunov exponent increases rapidly with the electric charge as $q$ approaches extremality. The behaviour observed in the eikonal limit, $\ell \rightarrow \infty$, where the spectra were computed analytically, agrees with the behaviour seen in figures (\ref{fig:modes})-(\ref{fig:modes3b}), where the frequencies were computed numerically for low angular degree, $\ell=1,2,3$. Therefore, we conclude that according to our findings, energy conditions imply a different pattern of the QN spectra.

\section{Conclusions}

To summarize our work, we have discussed here regular charged BH solutions within GR coupled to non-linear Electrodynamics. In particular, we have studied four models exhibiting distinct behaviors depending on the associated energy conditions. The scalar invariants related to the curvature of space-time as well as the electric field are finite at the origin, while the solutions are characterized by the mass, $M$, and the electric charge, $q$. The QNMs of scalar perturbations have been computed for all four models considered in this work, both numerically adopting the WKB approximation and analytically in the eikonal limit, $\ell \rightarrow \infty$. All modes are characterized by a negative imaginary part. The impact of the electric charge, the angular degree, $\ell$, and the overtone number, $n$, has been investigated in detail. The effective potential barrier has been shown graphically as a function of the radial coordinate, $r$, for all non-linear models for fixed $\{M, q\}$ and varying the angular degree. As far as the eikonal limit is concerned, both the critical frequency (real part of the frequencies) and the Lyapunov exponent (imaginary part of the frequencies) have been shown graphically as a function of the electric charge assuming that $M=1$ for all four models studied here. The critical frequency was found to be a monotonically increasing function of $q$, while the Lyapunov exponent was observed to exhibit a behavior that depends on the energy conditions associated to the model at hand. Besides, the numerical values of the frequencies that were numerically computed have been displayed in tables (\ref{table:First set})-(\ref{table:Fourth set}), while for better visualization the real part as well as the imaginary part have been shown graphically as a function of the electric charge assuming $M=1$ for different values of the angular degree and the overtone number.

\smallskip

In this article, we investigated regular black holes assuming non-linear electrodynamics in four-dimensional space-time. Also, we reviewed the importance of the energy conditions in various solutions. In addition, we have obtained the scalar quasinormal frequencies in all four cases, using i) the WKB semi-analytic approach and ii) in the eikonal limit ($\ell \gg 1$). Our results reveal that all modes are characterized by a negative imaginary part under massless and neutral perturbations. The real and imaginary part versus the electric charge exhibit different behaviour from one model to another depending on the energy conditions. In other words we find a correlation between energy conditions and pattern in the QN spectra. 
\section*{Acknowledgments}

The author A.~R. is funded by the Mar{\'i}a Zambrano contract ZAMBRANO 21-25 (Spain) (with funding from NextGenerationEU). The author L.~B. is supported by DIUFRO through the project: DI22-0026.






\end{document}